\documentstyle[psfig]{mn}

\def\etal{{\rm et al. }}
\def\mpc{{h^{-1} \rm Mpc}}

\title{The environment of active objects in the nearby universe}

\author[G. V. Coldwell, H. J. Mart\'{\i}nez \& D. G. Lambas]
       {Georgina V. Coldwell\thanks{On a fellowship from Agencia C\'ordoba
        Ciencia, C\'ordoba, Argentina},
        H\'ector J. Mart\'{\i}nez\thanks{CONICET fellow, Argentina} 
        and Diego G. Lambas\thanks{CONICET, Argentina}\\
        Grupo de Investigaciones en Astronom\'{\i}a Te\'orica y Experimental
        (IATE), Observatorio Astron\'omico,\\
        Universidad Nacional de C\'ordoba, Laprida 854, 5000, C\'ordoba, 
Argentina.\\
        e-mail: georgina@.oac.uncor.edu, julian@oac.uncor.edu, dgl@oac.uncor.edu}

\date{\today}

\begin{document}
\pagerange{\pageref{firstpage}--\pageref{lastpage}}

\maketitle

\label{firstpage}

\begin{abstract}
We study the galaxy environment of active galaxies, 
radio-loud and radio-quiet quasars in the redshift range
$0.1\leq z\leq0.25$. 
We use APM galaxies in order to explore the local galaxy overdensity 
and the $b_J-R$ colour distribution of neighbouring galaxies of these
target samples. 
For comparison, we perform similar analysis on samples of Abell 
clusters with X-ray emission, 
and samples of Abell clusters with richness  $R=1$ and $R=0$.
The projected cross-correlations show that the samples of quasars and active 
galaxies reside in regions of galaxy density enhancements
lower than those typical of $R=0$ clusters. 
We also find that in the nearby universe the local galaxy overdensity of 
radio-loud and radio-quiet quasars are comparable.
The analysis of the distribution of $b_J-R$
galaxy colour indexes suggests that the environment of quasars 
is not strongly dominated by a population of red galaxies,
characteristic of
rich Abell cluster 
, an effect that is more clearly appreciated for our sample of radio-loud quasars. 

\end{abstract}


\section{Introduction}
Studies of the environment of quasars have been the subject of several works 
in recent years. Some of the analysis in the local universe, $z< 0.6$,
suggest that quasars reside in groups (Fisher et al 1996), or 
in clusters (Smith et al 1999, Mc. Lure \& Dunlop 2000). However, 
Mart\'{\i}nez \etal (1999) using a quasar-galaxy cross-correlation analysis 
suggest that high density environments may not be 
representative of the typical quasar neighbourhood at low redshifts. 

At higher redshifts, $z>0.6$, some authors have found evidence of a different 
environment for radio-loud and radio-quiet quasars (Yee \& Green 1987,  
Yee 1990). Radio-loud quasars being associated mainly with
rich Abell clusters while radio-quiet quasars residing either
in groups or in the outskirts of clusters. However, more recent studies by 
Wold \etal (2000, 2001) and Best (2000) cast 
considerable doubts on previous results since
these new analysis do not show any appreciable difference between the typical
galaxy overdensity around radio-loud and radio-quiet quasars.
 
The discussion about the influence of quasar
activity on the formation and evolution of galaxies has continued
during the last decades.
The possibility that the surrounding galaxies could be 
responsible for triggering and fueling the quasars was first suggested 
by Toomre \& Toomre (1972). 
Interacting galaxies and mergers are efficient means of transporting gas
into the inner regions of the host galaxy. 
On the other hand, the large range of influence of a quasar
may have implications for structure formation. According to Rees (1988)
and Babul \& White (1991) quasars may ionize the surrounding medium 
up to several megaparsecs away. A different approach is adopted by other
authors (see for instance Voit 1996 and Silk \& Rees 1998),
where a fraction of  the energy released by quasars is transferred
via energetic outflows to the gaseous medium.

In this paper we attempt a characterization of the environment of quasars and
active galaxies in the nearby universe by considering the galaxy 
density enhancement using cross-correlation analysis with APM galaxies. 
We also use the distribution of $b_J-R$ colours of neighbouring galaxies
to explore their stellar population properties and deepen our understanding
of nuclear activity and galaxy formation and evolution. 
In section 2 we describe the data, section 3.1 deals with the cross-correlation
analysis, section 3.2 with the relative distribution of $b_J-R$ colour indexes 
of
APM galaxies around the different target samples. Finally, section 4 provides
a brief discussion of the main results.

\section{Data}

The active object target samples used in this work 
consist in radio-loud quasars, 
hereafter RLQs, radio-quiet quasars, hereafter RQQs, and active galaxies,
hereafter AGNs, from the 
Quasars and Active Galactic Nuclei 9nth Ed. Catalog 
(V\'eron-Cetty \& V\'eron 2000).  The cluster targets consist in 
clusters of galaxies taken from the X-ray Bright Abell Cluster Sample (XBACS)
(Ebeling \etal 1996) and Abell clusters with richness $R=1$ and $R=0$ 
selected from Abell \etal (1989).

All target samples have been restricted to the redshift range 
$0.1\leq z \leq 0.25$ taking into account the
similarity of the target redshift distributions 
within this range, and the large amplitude of correlation of the 
targets with APM galaxies.
Tracer galaxies within $0.5^{\circ}$ of the targets were taken from the 
APM Southern Sky Catalogues (http://www.ast.cam.ac.uk/\~{}apmcat/) and restricted 
to those targets for which data in the two bands $b_J$ and $R$ are available.
The final number of targets in each sample are listed in Table 1.
The samples of Abell clusters were restricted to richness $R\le1$
to avoid the inclusion of clusters already considered in the XBACS sample
since XBACS include mostly $R\geq 2$ clusters. 

\begin{table*}
\caption{Target sample characteristics.}
\centering
\begin{tabular}{lccc}\hline \hline
Sample & Number of objects & Source & Main characteristics \\
\hline
RQQs & 41 & V\'eron-Cetty \& V\'eron (2000) & $M_B\le-23$, radio-quiet\\
RLQs & 19 & V\'eron-Cetty \& V\'eron (2000) & $M_B\le-23$, radio-loud\\
AGNs & 35 & V\'eron-Cetty \& V\'eron (2000) & $M_B>-23$\\
XBACS & 34 & Ebeling \etal (1996)& $F_{lim}= 5.0\times10^{-12}$ 
erg cm$^{-2}$ s$^{-1}$\\
Abell $R=1$ & 35 & Abell \etal (1989) & $R=1$\\
Abell $R=0$ & 45 & Abell \etal (1989) & $R=0$\\
\hline
\end{tabular}
\end{table*}

We have also considered 30 randomly selected fields to make an appropriate
comparison with the mean background properties.
Main characteristics of the target samples analysed are also listed in Table 1.

\begin{figure}
\centerline{\psfig{file=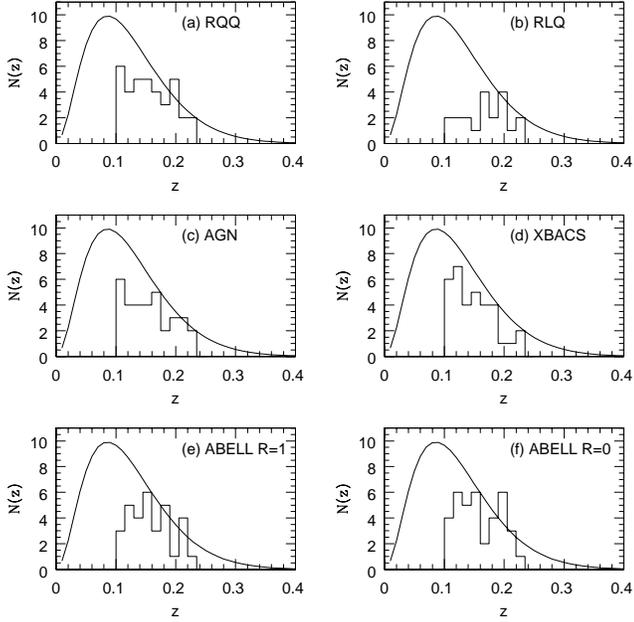,width=9cm,height=9cm}}
\caption{
Redshift distributions of the different samples.
The smooth curve represents the expected redshift distribution
for APM galaxies with limiting apparent magnitude $b_J^{{\rm lim}}=20.5$
as derived by Baugh \& Efstathiou (1993).}
\label{fig1}
\end{figure}

The different target samples have 
similar redshift distributions as it can be seen in
Figure 1, so that the results of our analysis are not subject to different
$K$-correction effects as well as other possible systematic that depend on
sample depth. For comparison, in each panel of Figure 1 is shown the
expected redshift distribution of APM tracer galaxies for a limiting
apparent magnitude $b_J^{{\rm lim}}=20.5$ (see Baugh \& Efstathiou 1993)
in arbitrary units. 

\section{Analysis}

\subsection{Galaxy overdensities around the targets}

In order to make a suitable characterisation of the typical galaxy density 
enhancement around the objects in the selected samples, we computed
the cross-correlation functions for centers of the different target samples
and APM tracers with a limiting magnitude $b_J^{\rm lim}=20.5$.
Since all center targets in our samples have measured redshifts, we are
able to compute cross-correlations as a function of projected distance
which improves the information obtained from angular correlations.

\begin{figure}
\centerline{\psfig{file=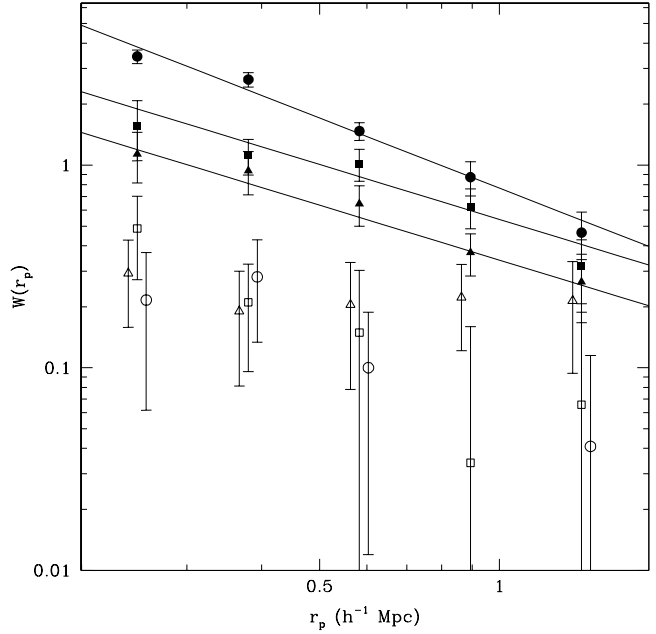,width=9cm,height=9cm}}
\caption{
Projected Cross-Correlation Functions. Filled circles correspond
to XBACS cluster sample, filled squares correspond to Abell $R=1$ and
filled triangles to Abell $R=0$. The best fit power-law to the projected
cross-correlation function for the cluster samples are shown (see parameters
in text). Empty squares correspond to RLQs, empty triangles to RQQs
and empty circles to AGNs.
}
\label{fig2}
\end{figure}

We have used the following estimator of the two-point correlation
function:
\begin{equation}
W(r_p)=\frac{DD(r_p)N_R}{DR(r_p)N_D}-1,
\end{equation}
where $DD(r_p)$ and $DR(r_p)$ are the number of target-galaxy 
and target-random point pairs respectively, separated in each
case by a projected distance $r_p$ in the interval 
$[r_p-\Delta r_p, r_p+\Delta r_p]$, $N_D$ and $N_R$ are 
the number of objects in the data and random catalogs respectively.  
When computing projected distances we have assumed a flat cosmology
($q_0=0.5$) and a Hubble constant $H_0=100h{\rm{kms^{-1}Mpc^{-1}}}$.
The resulting cross-correlation functions are shown in Figure 2. 
Error bars in these figures were estimated with the bootstrap 
resampling technique (Barrow et al. 1984) with 30 bootstrap samples.

In Figure 2 we show the projected cross-correlation functions corresponding
to the three cluster samples and the samples
of active objects, where it can be seen the decrease of the cross-correlation
amplitude moving from richer to poorer clusters as well as the lack of strong
correlation for the samples of active objects.
It is clear that the projected
cross-correlation functions of the samples of active objects have systematically
lower amplitude than those of cluster samples (even $R=0$) and they are indistinguishable
taking into account uncertainties. 

Assuming a power-law model for the spatial correlation function, 
$\xi(r)=(r/r_0)^{-\gamma}$, the projected correlation function is 

\begin{equation}
W(r_p)=C\pi^{1/2}\frac{\Gamma((\gamma-1)/2)}{\Gamma(\gamma/2)}
\frac{r_0^{\gamma}}{r_p^{(\gamma-1)}}.
\label{wrp}
\end{equation}
where the constant $C$, 
\begin{equation}
C=\frac{\sum_i p(y_i)}{\sum_i \left(\frac{1}{y_i^2} \right)
\int^\infty_0 p(x)x^2dx},
\label{c}
\end{equation}
takes into account the probability $p(x)$ for a tracer galaxy to be found
at a distance $x$ from the observer and is derivable from the galaxy luminosity
function $\phi(L)$:
\begin{equation}
p(x)\propto \int_{L_{\rm min}(x)}^{\infty}\phi(L)dL,
\end{equation}
where a galaxy at a distance $x$ must have a luminosity greater than 
$L_{\rm min}(x)$ to
be included in the tracer sample (see Lilje \& Efstathiou 1988).
In equation (\ref{c}) $y_i$ is the radial distance to the $i-$th center target. 
We have used a Schechter fit to the luminosity function (Schechter 1976)
for the tracer galaxies with parameters
$M^{\star}_{b_J}=-19.7 + 5\log h$, $\alpha=-1.0$ and we have included a 
linear $K$-correction, $K=3z$, in computing $L_{\rm min}(x)$.
By fitting power-laws to the cross-correlation functions
we have inferred slopes and correlation lengths for the cluster samples.
We obtain $\gamma=2.09\pm 0.09$, $r_0=(11\pm2)\mpc$ for XBACS;
$\gamma=1.9\pm 0.2$, $r_0=(11\pm3)\mpc$ for Abell $R=1$;
$\gamma=1.9\pm 0.2$, $r_0=(9\pm2)\mpc$ for Abell $R=0$. 

The low amplitude and the large error bars of the cross-correlations
of active objects do not allow adequate power-law fits. 
A useful quantity to characterise the typical galaxy overdensity
around the center targets considers the number of tracer 
galaxies at projected distances $r_p\le1\mpc$ with respect
to the random fields, 
\begin{equation}
N^{\rm excess}_{TARGET}=\langle N\rangle- N_{RAN}
\end{equation}
Our sample of tracer galaxies are limited to $b_J\le20.5$ and corresponds to
objects brighter than $M_{b_J}=-17.7 + 5\log h$ 
at $z\sim0.15$. We find for active objects
$N_{AGNs}^{{\rm excess}}\sim15$; $N_{RQQs}^{{\rm excess}}\sim13$;
$N_{RLQs}^{{\rm excess}}\sim10$. For clusters we find 
$N_{XBACS}^{{\rm excess}}\sim100$; 
$N_{R=1}^{{\rm excess}}\sim50$ and $N_{R=0}^{{\rm excess}}\sim30$. 
From these results it is clear that active objects are not typically
located in moderately rich galaxy enhancements such as Abell $R=0$ clusters,
as suggested in previous works (see for instance Smith \etal 1999,
Mc. Lure \& Dunlop 2000).

\subsection{Distribution of $b_J-R$ colour indexes of neighbouring galaxies}

The relative fraction of different stellar populations in galaxies
are reflected in the $b_J-R$ colour index. 
Elliptical galaxies, dominated by an old population
of stars, have typical values $b_J-R \sim 1.5 - 2$, while in late types 
with recent episodes of star formation, $b_J-R < 1$.
In this subsection we explore the properties of the 
stellar populations of galaxies in the vicinity of the different target
samples using $b_J-R$ colour indexes.
We study the normalised distribution of $b_J-R$ 
of APM galaxies around the targets within a projected distance
$r_p< 0.5 \mpc$. We adopt  the same limiting 
apparent magnitude $b_J^{{\rm lim}}=20.5$ for all tracer samples and 
we compute:  
\begin{equation}
f(b_J-R)=\frac{N(b_J-R)}{N^{TOT}\Delta (b_J-R)},
\end{equation}
where $N(b_J-R)$ is the number of APM galaxies 
with colour indexes in the interval defined by 
$[(b_J-R)-\Delta (b_J-R)/2, (b_J-R)+ \Delta (b_J-R)/2]$, 
and $N^{TOT}$ is the total number of APM galaxies within $r_P\le 0.5\mpc$.
Error bars in these distributions were computed using the 
bootstrap resampling technique. 
We subtract from these distributions the colour 
distribution of tracer galaxies in the randomly selected fields. 
The results are displayed in Figure 3.  As it is expected from the well known
morphology-density relation (Dressler 1980), there is a marked relative excess of 
early ($1\leq b_J-R \leq 2$) to late type galaxies 
($0\leq b_J-R\leq 1$) within $r_p\leq 0.5\mpc$ for the three samples 
of clusters, an effect that is stronger for the richer cluster samples.
Active objects show a weaker tendency, although, 
the distribution of colours of RQQs
shows a similar excess of red galaxies than in $R=0$ clusters. 
For RLQs, the signal for the presence of a red galaxy population is almost negligible,
while AGNs show an intermediate behavior between  RQQs and RLQs. 

\begin{figure}
\centerline{\psfig{file=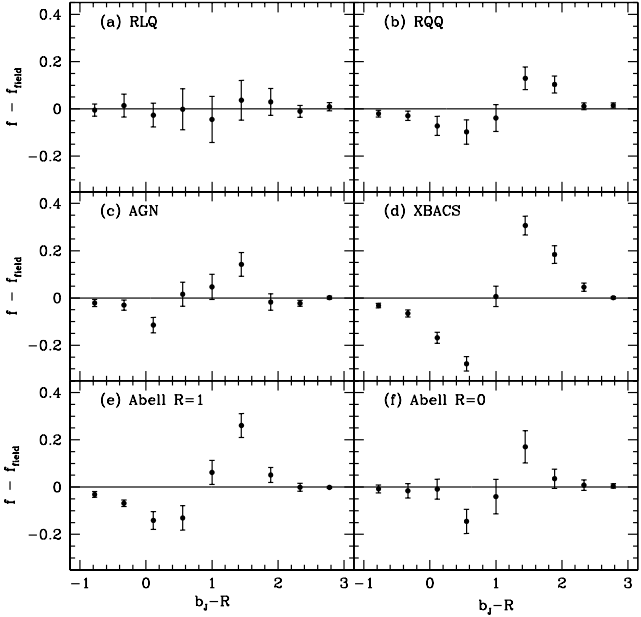,width=9cm,height=9cm}}
\caption{
Colour index $b_J-R$ distributions
normalised to randomly selected fields corresponding to
galaxies within $r_p\le0.5\mpc$ from the targets.
}
\label{fig3}
\end{figure}

To quantify the previous results, we have computed a parameter $\cal P$ 
that characterizes the relative fraction of red to 
blue galaxies in a given range of projected separations

\begin{table*}
\centering
\caption{
${\cal P}$ parameter for the different target 
samples within three projected distances. 
}
\begin{tabular}{lccc}
\hline
\hline
Sample & ${\cal P}$& ${\cal P}$ & ${\cal P}$ \\
 & $r_p\le 0.3\mpc$ & $r_p\le 0.5\mpc$ & $r_p \le1\mpc$ \\
\hline
RQQs         & $ 0.10\pm0.06$ & $ 0.09\pm0.04$ & $ 0.07\pm0.02$\\
RLQs         & $-0.22\pm0.08$ & $-0.05\pm0.07$ & $-0.05\pm0.04$\\
AGNs         & $ 0.08\pm0.05$ & $ 0.03\pm0.04$ & $-0.01\pm0.02$\\
XBACS       & $ 0.49\pm0.03$ & $ 0.38\pm0.02$ & $ 0.30\pm0.02$\\
Abell $R=1$ & $ 0.17\pm0.05$ & $ 0.20\pm0.04$ & $ 0.16\pm0.02$\\
Abell $R=0$ & $ 0.18\pm0.07$ & $ 0.09\pm0.04$ & $ 0.06\pm0.02$\\
\hline
Field   & $-0.11\pm0.01$\\
\hline
\end{tabular}
\end{table*}

\begin{equation}
{\cal P}=\frac{N_R-N_{b_J}}{N_R+N_{b_J}},
\end{equation}
where $N_R$ is the number of neighbour tracer galaxies in the range 
$1\leq b_J-R \leq 2$ and $N_{b_J}$
is the number in the range $0\leq b_J-R \leq 1$. In Table 2 we list this 
parameter for three projected distance ranges 
($r_p\le0.3\mpc$, $ r_p\le 0.5\mpc$ and $ r_p\le 1\mpc$) and their
corresponding bootstrap resampling uncertainties estimates. 

The values of ${\cal P}$ in Table 2 allow us to quantify and extend
the previous results of this subsection. By inspection to this table
it can be seen that  galaxies in clusters show the tendency of 
systematically larger colour indexes near the center.    
The population of galaxies in the neighbourhood of RQQs 
have ${\cal P}$ comparable to that of galaxies in $R=1$ and $R=0$ Abell clusters.
It is worth to notice the negative values of ${\cal P}$ for
galaxies in the vicinity of RLQs which indicate the lack of a red galaxy
population, characteristic of high density regions, around these objects.
In order test the stability of these results we have changed the boundaries 
adopted to calculate the values of ${\cal P}$. The analysis shows that the 
above conclusions are robust and do not depend on the particular choice of
colour boudaries in the calculations, provided the
parameter ${\cal P}$ reflects the relative fraction of early-type to late-type
galaxies.  

\section{Discussion}

The results discussed previously lead us to
conclude that  quasars and AGNs are not placed in 
galaxy density enhancements corresponding to rich and moderately rich clusters 
of galaxies. 
As it can be clearly appreciated by inspection to Figure 2, 
the galaxy overdensity around these center targets is
significantly lower than that corresponding
to Abell $R=1$ or $R=0$ clusters of galaxies. 
Thus, active objects are more likely to reside in groups
(typically 10 neighbours with $M_{b_J}\le M^{\star}_{b_J}+2$ 
within $r_p\le1\mpc$).
It has been often argued that quasar environment correspond to that of
moderately rich clusters, our results indicate that 
active nuclei reside in regions of at most half the projected galaxy
density of $R=0$ clusters.

The distribution of $b_J-R$ colour indexes of galaxies within projected
distances $r_p\le 1 \mpc$ centered on RLQs is significantly 
different than that corresponding to clusters.
This fact suggests that the stellar population of the neighbouring 
galaxies of RLQs differs significantly from that in rich and
moderate galaxy density enhancements. 
The environments of RQQs show $b_J-R$
colour indexes similar to galaxies in clusters while AGNs show an 
intermediate behavior.

We find statistical evidence that the distribution of $b_J-R$ colours of
galaxies in the vicinity of radio-quiet and radio-loud quasars are different. 
The latter have a population of neighbouring galaxies with lower values of
colour indexes. This fact should be considered in models of the 
effects of nuclear activity on galaxy formation and evolution.

\section{Acknowledgments}
We thank Dr. Howard Yee for useful comments. 
We are indebted to Dr. Carlton Baugh for reading the 
manuscript and providing helpful suggestions. The authors would like
to thank the anonymous referee for helpful comments that improved 
the previous version of this paper.

This research was partially supported by grants from  CONICET,
Agencia C\'ordoba Ciencia, Secretar\'{\i}a de Ciencia y T\'ecnica
de la Universidad Nacional de C\'ordoba and Fundaci\'on Antorchas, Argentina.

\label{lastpage}
\end{document}